\pdfoutput=1
\documentclass[conference]{IEEEtran}
\usepackage{xcolor, soul}
\sethlcolor{yellow}
\usepackage{amsmath} 
\usepackage{graphicx}
\usepackage{multirow}
\usepackage{lipsum}

\usepackage[ruled,vlined]{algorithm2e}

\usepackage{graphicx}
\usepackage{textcomp}
\usepackage{tcolorbox,xcolor}
\tcbuselibrary{theorems}
\newtcbtheorem{mytheo}{Procedure}%
{colback=black!5,colframe=red!50!black,fonttitle=\bfseries}{}
\begin{document}
\title{A Challenge Obfuscating Interface for
       Arbiter PUF Variants against Machine Learning Attacks}

\author{\IEEEauthorblockN{Yu Zhuang}
\IEEEauthorblockA{
Department of Computer Science\\Texas Tech University\\ Lubbock, TX 79409, USA\\Email: Yu.Zhuang@ttu.edu}
\and
\IEEEauthorblockN{Khalid T. Mursi}
\IEEEauthorblockA{College of Computer Science and Engineering\\ University of Jeddah\\ Jeddah 21959, Saudi Arabia\\Email: kmursi@uj.edu.sa}
\\
 March 2021
\and
\IEEEauthorblockN{Li Gaoxiang}
\IEEEauthorblockA{Department of Computer Science\\Texas Tech University\\ Lubbock, TX 79409, USA\\Email: gaoli@ttu.edu}

}

\maketitle

\begin{abstract}
Security is of critical importance for the Internet of Things (IoT). Many IoT
devices are resource-constrained, calling for lightweight security protocols.
Physical unclonable functions (PUFs) leverage integrated circuits' variations 
to produce responses unique for individual devices, and hence are not reproducible
even by the manufacturers. Implementable
with simplistic circuits of thousands of transistors and operable with low energy, 
Physical unclonable functions are promising candidates as security primitives for 
resource-constrained IoT devices. Arbiter PUFs (APUFs) are a group of delay-based
PUFs which are highly lightweight in resource requirements but suffer from high
susceptibility to machine learning attacks.  To defend APUF variants against 
machine learning attacks, we introduce challenge input interface, which incurs 
low resource overhead. With the interface, experimental attack study shows that 
all tested PUFs have substantially improved their resistance against machine 
learning attacks, rendering interfaced APUF variants promising candidates for
security critical applications. 
\end{abstract}


\IEEEpeerreviewmaketitle

\section{Introduction}

Security is critically important for the Internet of Things (IoT) \cite{miorandi2012internet}. 
Key-based cryptographic protocols are popular choices for existing security applications. 
With conventional cryptographic protocols, there still remain two security challenges. 
First, many IoT devices are resource-constrained and support only a low level of 
         operating power, making existing cryptographic protocols unsuitable, since, according 
         to studies \cite{herder2014physical, yu2016lockdown}, they are not lightweight enough 
         in resources requirement. 
Second, cryptographic keys have to be stored in non-volatile memories. But many IoT
         devices are within close physical distances to the crowd, unlike data center servers
         which are locked in high-security rooms inaccessible to the crowd. Close physical 
         distances make these IoT devices vulnerable even with security protocols, since any 
         data stored in nonvolatile memories, including secret keys, can be exposed by 
         side-channel attacks 
         \cite{kommerling1999design, skorobogatov2005semi, yarom2014flush+, ruhrmair2014pufs}, 
         which are usually more effective within close distances.

Physical unclonable functions (PUFs) are an emerging class of hardware primitives for 
implementing security protocols. Small scale variations of integrated circuits exist in 
silicon chips. These variations are regarded as side effects for conventional integrated 
circuits \cite{suh2007physical}, but they make each chip unique and can be exploited to 
prevent semiconductor re-fabrication. PUFs utilize such variations to produce responses 
unique for individual PUF circuits 
\cite{herder2014physical,yu2016lockdown, ruhrmair2014pufs, 
      gassend2002controlled,gassend2002silicon,suh2007physical} 
and hence are not reproducible even by the PUF manufacturers. These physical variations are 
hardware fingerprints that can be used for security purposes. Such features of PUFs present 
potential solutions for the two challenges.

Instead of storing secrets (e.g. cryptographic keys) in nonvolatile memories, PUFs 
      retrieve the secret information present in the inevitable physical variations to 
      produce unique responses as signatures of silicon chips. With physical variations 
      occurring at multiple locations of the circuit (e.g. different delays of different gates), 
      some PUFs can create exponentially many combinations of these variations to produce 
      circuit-dependent responses, 
      leading to exponentially many secret keys. The huge number of keys enables each key
      to be used only once, promising an unlimited supply of one-time keys, namely keys that
      do not have to be stored in devices' memories and have very low risks even when exposed.
Implementable with simplistic circuits with only thousands of transistors, PUFs require 
      low fabrication cost and consume very low operating energy, rendering them excellent 
      candidates for resource-constrained IoT devices.

Though physically unclonable, some PUFs are ``mathematically clonable'' in the sense that the 
responses of a PUF can be predicted accurately by machine-learning methods. Attackers can 
eavesdrop on the communications between a PUF and its trusted partner, and the challenges 
sent to a PUF and the responses from the PUF to the challenges which are used in communication 
authentication can be collected by attackers to train machine learning models. Such models 
can accurately predict the responses of the PUFs after the models are trained with sufficient 
challenge-response pairs (CRPs). When equipped with openly accessible interfaces, PUFs are 
even more vulnerable when each challenge can be repeatedly applied to a PUF to launch 
reliability-based machine learning attacks \cite{becker2015gap}. Such mathematical 
clonability allows adversaries to develop malicious software to launch spoofing attacks.

Thus, before being deployed in real applications, a PUF must be examined to identify all 
possible security vulnerabilities. Detected vulnerabilities will be useful to both application 
developers and PUF designers. Such information can guide PUF designers to develop new PUFs to 
overcome existing vulnerabilities, and IoT application developers can use the information of 
detected PUF vulnerabilities to avoid PUFs whose vulnerabilities cannot be eliminated at the 
application level, and use other PUFs with no risks or different risks that can be taken care 
of with application-specific techniques. 

The arbiter PUF \cite{} and its variant are delay-based strong PUFs, which admit 
exponentially many challenge-response pairs (CRPs). The arbiter PUF and its variants, like
XOR PUFs \cite{suh2007physical}, Lightweight Scure PUFs (LSPUFs) \cite{majzoobi2008lightweight},
FF PUFs \cite{gassend2004identification,lee2004technique, lim2004extracting},  
multiplexer PUFs \cite{sahoo2017multiplexer}, and 
Interpose PUFs \cite{nguyen2019interpose}, are of low circuit complexity and hence highly 
lightwieght in demanding implementation resources and operation power. But studies showed 
that they suffer from susceptibility to machine learning attacks
\cite{ruhrmair2010modeling}, and the vulnearble ones include  
XOR PUFs \cite{tobisch2015scaling,aseeri2018XOR,electronics9101715},  multiplexer PUFs 
\cite{santikellur2019deep,mursi2020MPUF_JCM}, LSPUFs with 6 or fewer component arbiter PUFs 
\cite{santikellur2019deep}, FF PUFs \cite{ruhrmair2013puf}, and Interpose PUFs 
\cite{santikellur2019deep,wisiol2020splitting}.

\begin{figure*}[]
  \centering
  \includegraphics[width = 5in]{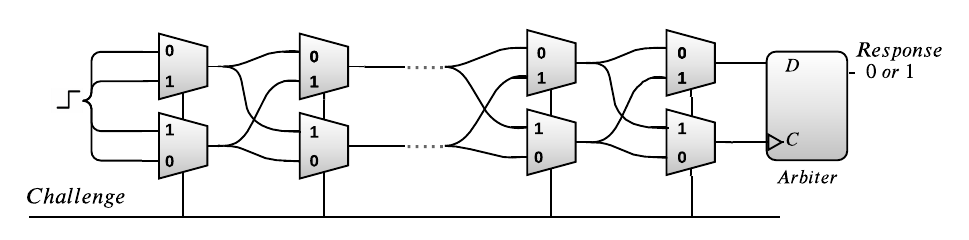}
  \caption{An arbiter PUF}
  \label{fig:APUF}
 \end{figure*}
 
 \begin{figure*}[]
  \centering
  \includegraphics[width = 5in]{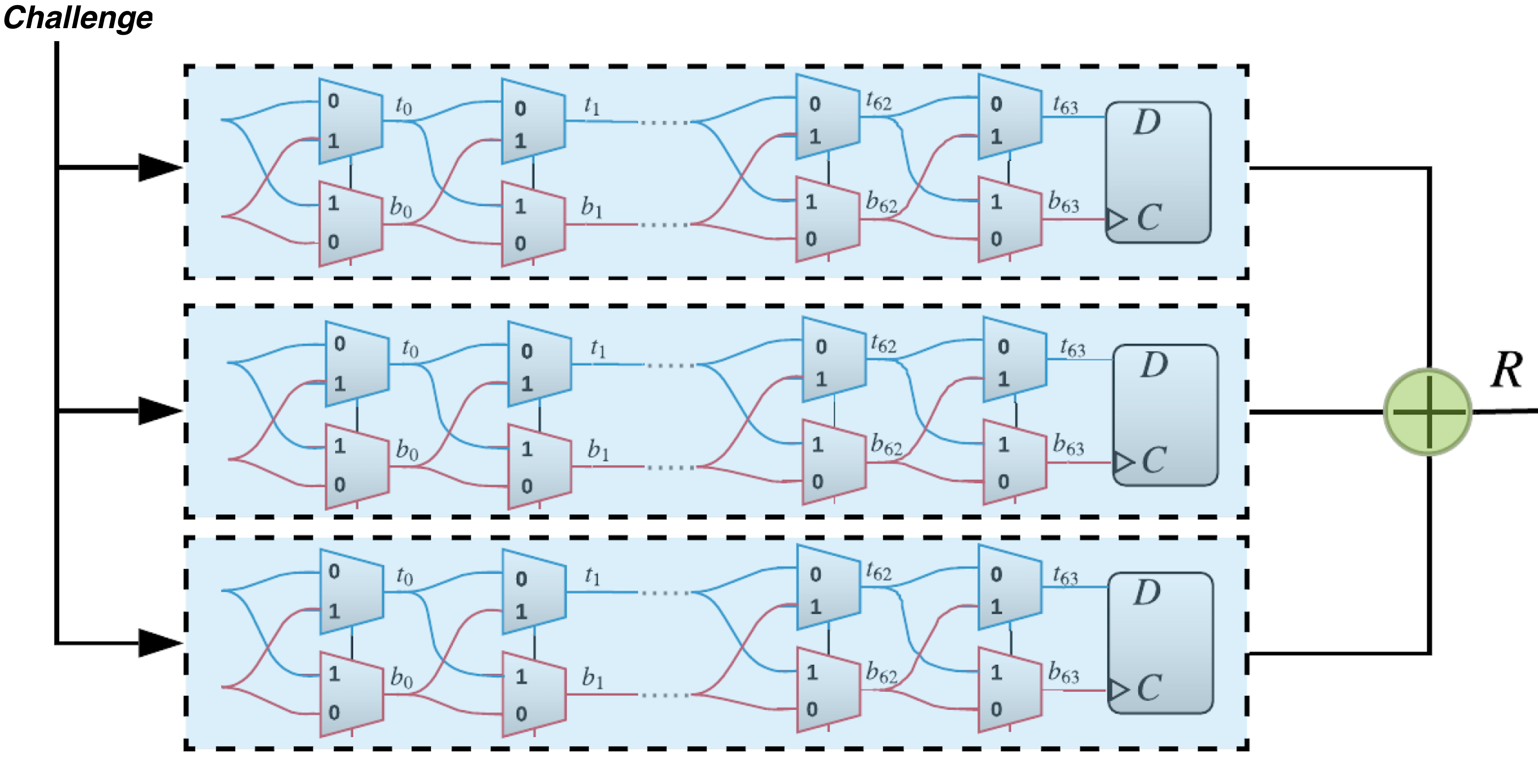}
  \caption{An XOR arbiter PUF}
  \label{fig:Xor_PUF}
 \end{figure*}
 
\begin{figure*}[]
  \centering
  \includegraphics[width = 5in]{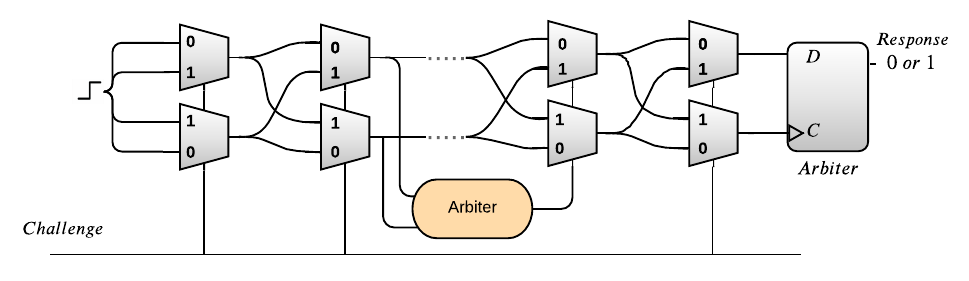}
  \caption{A feed-forward arbiter PUF with one feed-forward loop}
  \label{fig:FFPUF}
 \end{figure*}



The vulnearbility of PUFs to machine learning attacks has motivated us to
look for defensive methods, having led us to a challenge input interface. The interface 
uses ghost input bits which are not fed to the PUF as challenge bits, and the ghost bit 
positions are randomized for different PUF instances to obfuscate the challenge. 
The defense interface works under the threat model that attackers cannot choose challenges
but can obtain CRPs only through eavesdropping on the communications between the 
PUF and its trusted partner, which exlcuding the possibility of applying any
challenge repeatedly to launch the reliability attack \cite{becker2015gap}.  This 
threat model is achieveable since the lockdown protocol \cite{yu2016lockdown} can be 
employed to prevent attackers from choosing challenges.

Our experimental study showed that with this interface, 
\begin{itemize}
    \item arbiter PUFs, XOR PUFs, and FF PUFs have substantially increased their
          security against machine learning attacks, and 
    \item all 64-bit APUFs, XPUFs, and FFPUFs can withstand our 
          neural network attacks when enough ghost bits are used by the interface.
\end{itemize}

\section{The Arbiter PUF and its Variants}

For clarity of technical discussions in later sections, in the following we give a brief 
description of the circuit architectures of the Arbiter PUF and its variants.  

\subsection{The Arbiter PUF}
An $n$-stage arbiter PUF consists of $n$ pairs of 
2-to-1 multiplexers (see Fig. \ref{fig:APUF}), where the two multiplexers at the same stage 
receive the same challenge bit. The two signals pass through gates of all stages of the paths, 
and slightly different delays are incurred when signals pass through different gates. An 
arbiter, usually implemented by a D-latch, determines the final output depending on which 
signal arrives first. For instance, if the top path arrives first, the output is 1, otherwise 
is 0. The challenge bit values at all stages determine the paths, and consequently the delays 
of the signals, leading to a total of $2^n$ possible paths.

The arbiter PUFs satisfy the additive delay model \cite{lim2004extracting}, which stipulates 
that the time it takes for each of the two signals to arrive at the arbiter are the summation 
of the delays incurred at all stages of the PUF. Based on the additive delay model, the 
response $r$ of an arbiter satisfies
\begin{equation}
\label{equ:one}
r = Sgn(v(n) + \sum_{i=1}^{n} w(i)\phi(i)) ,
\end{equation}
where $\phi$'s are transformed challenge \cite{lim2004extracting} given by
\begin{equation}
\label{equ:two}
\phi(i) = (2c_i-1)(2c_{i+1}-1)\cdot\cdot\cdot\cdot\cdot(2c_n-1),
\end{equation}
with $c_i$ being the challenge bit at stage $i$, $v$ and $w$'s being parameters quantifying 
gate delays at different stages, and $Sgn(\cdot)$ the sign function.

In (\ref{equ:one}), the term inside the $Sgn(\cdot)$ function is linear with respect to the 
transformed challenge $\phi$'s. The model represented by (\ref{equ:one}) is hence a linear 
classification problem with the separating hyperplane represented by equation
\begin{equation}
\nonumber
w(1)\phi(1)+w(2)\phi(2)+\cdot\cdot\cdot\cdot\cdot+w(n)\phi(n)+v(n)=0,
\end{equation}
which results from setting to 0 the term inside the $Sgn(\cdot)$ function in (\ref{equ:one}). 
The hyperplane is in the $n$-dimensional space of transformed challenges, which partitions 
the space into two regions, with 1 as the response to transformed challenges in one region 
and -1 as the response to transformed challenges in the other region. For a machine learning 
attack method that has accumulated enough CRPs, the CRPs are used to train the machine 
learning model so that the trained model can accurately predict the responses to other 
challenges. The linear classification model (\ref{equ:one}) describing the relationship 
between the response $r$ and the transformed challenge $\phi$ makes arbiter PUFs vulnerable 
to machine learning attacks \cite{ruhrmair2010modeling}.

\subsection{The XOR Arbiter PUF}
An XOR arbiter PUFs \cite{ruhrmair2010modeling, suh2007physical}, or XOR PUFs for short,  
is built from multiple arbiter PUFs. As illustrated in Fig. \ref{fig:Xor_PUF}, the $k$-XOR 
arbiter PUF uses $k$ arbiter PUFs as components, where all of the $k$ arbiter PUFs use the 
same challenge $C$ as the challenge input. The responses of all individual arbiter PUFs 
are XORed together to produce the final response $r$ for the corresponding input challenge 
$C$. Thus, the response of the $n$-stage $k$-XOR arbiter PUF in Figure \ref{fig:Xor_PUF} can 
be expressed as:

\begin{equation}
\textbf{\emph{r}} = \mathop{\bigoplus}\limits_{j = 1 \ldots k} r_{j},
\label{eq_3}
\end{equation} 

where $r_{j}$ is the internal output of the $j^{th}$ component arbiter PUF.

The XOR operation increases non-linearity of the relationship between the response $r$ and 
the transformed challenges $\phi$'s. It is obvious that adding arbiter PUFs increases the 
cost needed for silicon implementation of the PUFs. But, every 
additional arbiter PUF increases nonlinearity as well as the dimensionality of the 
parameter space to be machine-learned by attackers \cite{ruhrmair2010modeling}, leading 
to higher resistance against machine learning attacks \cite{lim2005extracting}. For more 
details on PUF designs, the types, and their specific variations, we refer the reader to 
\cite{ruhrmair2010modeling,herder2014physical}.

\subsection{The Feed-Forward Arbiter PUF}
%

The Feed-Forward Arbiter PUF (FF PUF) 
\cite{gassend2004identification, lee2004technique, lim2004extracting} 
was introduced 
to increase the complexity of the relationship between the challenge and response. An FF~PUF 
resembles an arbiter PUF but with some of the challenge input bits taking the outputs of 
feed-forward arbiters, with each feed-forward arbiter receiving the output bits of the two 
multiplexers at an earlier stage of the PUF. Fig. \ref{fig:FFPUF} is an illustration of an 
FF PUF with one feed-forward loop. The use of internal signals as input challenge bits to 
multiplexers makes the function relationship between the response and the input challenge 
more complex than that of the arbiter PUF.

By adding feed-forward loops, the challenge input to an FF PUF at a loop-ending stage is the 
arbiter output of signal delay difference at an earlier stage, which can be modeled by an 
equation similar to (\ref{equ:one}).  For a one-loop feed-forward PUF with $n$ stages of 
multiplexers, if the loop starts at stage $i_1$ and ends at stage $i_2$ (assuming $i_2<n$), 
then based on model (\ref{equ:one}) for arbiter PUFs, the response of a single-loop FF PUF 
can be modeled by
\begin{equation}
\label{equ:three}
r=Sgn( v(n)+\sum_{i\neq i_2} w(i)\phi(i ) + w(i_2 )\phi(i_2 )),
\end{equation}
where
\begin{equation}
\nonumber
\begin{cases}
\phi(n)=2c_n-1,\\
\phi(i)=(2c_i-1)\phi(i+1)    \text{ for  } i \neq i_2  \text{ and i} < n\\
\phi(i_2) = Sgn(v(i_1) + \sum_{i=1}^{i_1} w(i)\phi(i)) \text{ with}\\
\phi(i_1) = 2c_{i_1}-1, \phi(i)  = (2c_i‒1)\phi(i+1)  \text{ for } i<i_1.
\end{cases}
\end{equation}

Since the sign function $Sgn(\cdot)$ is nonlinear and non-differentiable, the transformed 
challenge bit $\phi(i_2)$ at the loop-ending stage is a nonlinear non-differentiable function 
of challenge bits of earlier stages. This nonlinearity and non-differentiability turn the 
linearly separating hyperplane for the arbiter PUF into a nonlinear separating surface for 
the FF PUF. Since machine learning methods are in general less effective for modeling 
nonlinearly separated classes than linearly separated ones, the FF PUFs are more resistant 
against machine learning attacks.

\section{Defensively Interfaced Strong PUFs}

To defend PUFs against machine learning attack, we propose an interface for 
the input challenge which, as elaborated in the following bullets, requires the trusted 
partner of the PUF to send more bits than the number of challenge bits to the PUF. 
\medskip

\noindent{\bf ~ The $m$+bits Interface}

\begin{itemize}
\item For a PUF with $n$-bit challenge, the challenge input interface has $(n+m)$ 
      input bits. Input bits are labeled as $b_1,b_2,\cdots,b_n, \cdots, b_{n+m}$.
\item A random selection of $n$ input bits out of $(n+m)$ is chosen for each PUF instance. 
      The $n$ bits are supplied to the PUF instance as the challenge bits, and remaining 
      $m$ input bits, called ghost bits, are not used by the PUF.
\end{itemize}

Since the defensive interface selects a random subset of input bits as challenge bits for
each PUF instance, attackers have no knowledge of the which $n$ bits of the $(n+m)$ input 
bits are used for the challenge and which input bits are not used, leading to an obfuscation
of the challenge. But a trusted partner of the PUF, say a securely protected server partner 
of the PUF, knows which of the input bits are challenge bits and will be able to generate 
the response using a PUF model stored in the trusted partner to verify the PUF. 

Also, we assume that attackers cannot use cannot choose 
challenges and apply them to the PUF to get 
responses, but can obtain CRPs only through eavesdropping on the communications between the 
PUF and its trusted partner.  Without this assumption, an attacker can guess which bits of 
the challenge are not used by flipping one bit at a time but keeping all other bits of 
challenges fixed, and if the responses do not change, then the flipped bit is not used. 
But with this assumption, this type of guessing attacks cannot be employed. This assumption 
can be realized when the lockdown protocol \cite{yu2016lockdown} is employed.

It is obvious that this challenge-obfuscation interface incurs low resource overhead. To the 
best of our knowledge, all existing CRP obfuscating protocols require hardware resources for 
their implementations and require operations to perform the obfuscation, e.g. protocols 
\cite{gao2017puf, majzoobi2008lightweight, rostami2014robust, ye2016rpuf} requires TRNG and 
additional transistors to make use of TRNG, and fuzzy extractor is used in \cite{van2012reverse}.
 Our proposed interface requires only a small number of additional bits for the input and does 
not use any transistors, and is hence of low circuit overhead and low energy overhead.

\section{Experimental Attack Study of Interfaces PUFs}

Earlier attack studies on PUFs 
\cite{ruhrmair2010modeling,aseeri2018XOR,santikellur2019deep,electronics9101715} 
show that neural networks have high modeling power of PUF bahavior. 
To experimentally examine the effectiveness of our proposed interface against modeling attacks, 
we apply neural network attack methods on arbiter PUFs, XOR PUFs, FF~PUFs,and their interfaced
counterparts.

\subsection{Attack Study on APUFs, XPUFs, and Their Interfaced Counterparts}

To examine the effectiveness of the defensive interface for arbiter PUFs and XOR PUFs, we applied
the same neural network attack method to arbiter PUFs and interfaced arbiter PUFs, and applied 
the neural network attack method to 3-XOR PUFs and interfaced 3XOR PUFs. All defensive interfaces
use 16 ghost bits.
%
%
The neural networks we used in the attack study follow the specification given in
\cite{electronics9101715}, and for the convenience of reading are listed in
Table \ref{tab1}, where arbiter PUF is denoted by $k$-XPUF with $k\!=\!1$. 
We implemented machine learning methods in Python code using 
the Keras machine learning library \cite{NEURIPS2019_9015}.

\def\arraystretch{1.5}
\begin{table}[h] 
\caption{Parameters of the Neural Network for attacking $k$-XPUF}
\centering
\begin{tabular}{|l|c|}
\hline
\multicolumn{1}
   {|c|}{{\ul \textbf{Parameter}}}  & {\ul \textbf{Description}} \\    \hline
   \textbf{Optimizing Method}       & ADAM                       \\    \hline
   \textbf{Hidden Lyr. Actv. Fx.}   & tanh                       \\    \hline
   \textbf{Output Lyr. Actv. Fx.}   & Sigmoid                    \\    \hline
   \textbf{Learning Rate}           & Adaptive                   \\    \hline
   \textbf{Network Size}            & 3 hidden lyrs. ($32, 64, 32$) \\    \hline
   \textbf{Loss Function}           & Binary cross entropy       \\    \hline
   \textbf{Mini-batch Size}         & 1000                       \\    \hline
   \textbf{Patience}         & 5                       \\    \hline
\end{tabular}
\label{tab1}
\end{table}

In the tests, we generated 30 simulated arbiter PUF instances using the simulator
pypuf \cite{nils_wisiol_2020_3904267} from the normal distribution with a mean of 0 and a standard deviation of 1, plus an noisiness value 0.01 for the generated PUFs. 
For each PUF instance 5 million CRPs are generated .    
With the CRPs ready, we carried out tests of the attack on a PC with a 3.6 GHz 
6-Core AMD Ryzen 5 3600 processor and a memory capacity of 16 GB. Each machine learning method
for each simulated PUF instance is optimized for up to 300 epochs with an early stopping 
of 5 patience. The experimental study used an 85-5-10 training-validation-testing split, 
where 85\% of CRP data were used for training, 5\% of data were used for validation, and 
10\% of data for testing the model. In each attack, the number of CRPs used in the attack 
started small and gradually increasing until having reached a size (the size listed in the 
Column CRPs in Table \ref{tab3}), which resulted in successful attacks for all 30 PUF instances. 
The experimental results of the attacks on are listed in Table~\ref{tab3}.

\begin{table}[]
\centering
\caption{Results of Attacks on 64-stage $k$-XPUFs with and without 16+bits Interface}
\label{tab3}
\begin{tabular}{|c|c|c|c|}
\hline
{\textbf{PUF Type}}&\textbf{CRPs}&\textbf{Accuracy}&\textbf{Time}\\ \hline
1-XPUF                               &  5 K        & 98 \%           & 0.9 sec          \\ \hline
3-XPUF                              &  9 K        & 98 \%           & 2.8 sec        \\ \hline
Interfaced 1-XPUF                   &  4.5M       & 66 \%           & No convergence          \\ \hline
Interfaced 3-XPUF                   &  4.5M       & 51 \%           & No convergence          \\ \hline
\end{tabular}
\end{table}

%
%
%
%

\subsection{Attack Study on FF~PUFs and Interfaced FF~PUFs}

The attack model for our attack method is that the attacker has no knowledge of the 
internal circuit structure of the PUF, but have accumulated enough CRPs where each CRP of 
an $n$-stage $k$-loop FF PUF has ($n-k$) challenge bits and one response bit. Our 
attack model is the same as the one used for FF PUFs in the study \cite{ruhrmair2010modeling}
though the unavailability of internal PUF circuit information is not explicitly stated in 
\cite{ruhrmair2010modeling}.

To attack FF~PUFs, we employed a neural network method. We tried different number of layers
and different number of neurons for different layers, and the following network architecture
is the one we used in the attack study.

\begin{figure}[t]
 	\centering
 	\includegraphics[width = 3.5in]{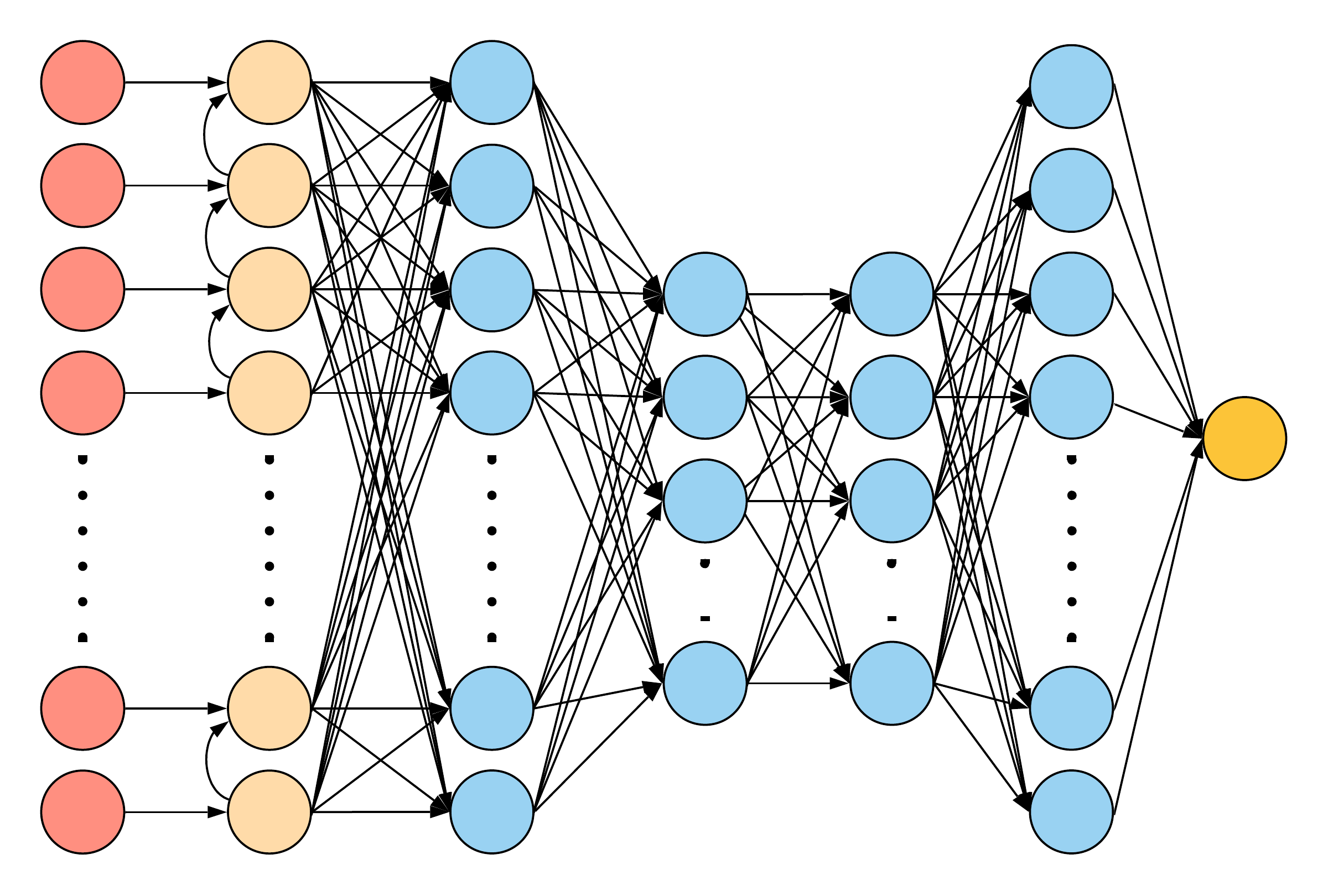}
 	\caption{Illustration of the neural network used in attacking FF~PUFs.}
 	\label{fig:FF_NN.png}
 \end{figure}

The neural network we use for modeling an $n$-stage $k$-loop FF PUF has the following 
structure of layers (also see Fig. \ref{fig:FF_NN.png} 
for illustration):
\begin{itemize}
\item The input layer of ($n‒k$) challenge bits $c_1,c_2, ..., c_{n-k}$,
\item the challenge transform layer that transforms challenge bits from input layer to 
      $\phi$'s according to \\
      $\phi(i) = (2c_i-1)(2c_{i+1}-1)\cdot\cdot\cdot(2c_{n-k}-1),$
\item four layers of the densely connected neural network whose weights for all neurons 
      are to be trained, and
\item the single-bit output layer to produce the output that models the response of the 
      FF PUF.
\end{itemize}

Other parameters of the neural network for attacking FF PUFs are listed in Table \ref{tab:mlp_parameters}.

\def\arraystretch{1.5}
\begin{table}[h] 
\caption{Parameters of the Neural Network}
\centering
\begin{tabular}{|l|c|}
\hline
\multicolumn{1}
   {|c|}{{\ul \textbf{Parameter}}}  & {\ul \textbf{Description}} \\    \hline
   \textbf{Optimizing Method}       & ADAM                       \\    \hline
   \textbf{Hidden Lyr. Actv. Fx.}   & tanh                       \\    \hline
   \textbf{Output Lyr. Actv. Fx.}   & Sigmoid                    \\    \hline
   \textbf{Learning Rate}           & Adaptive                   \\    \hline
   \textbf{Network Size}            & 4 hidden lyrs. (64, 32, 32, 64) \\    \hline
   \textbf{Loss Function}           & Binary cross entropy       \\    \hline
   \textbf{Mini-batch Size}         & 3000                        \\    \hline
   \textbf{Patience}         & 5                       \\    \hline
\end{tabular}
\label{tab:mlp_parameters}
\end{table}

The CRPs we used for the attack were generated by simulated FF PUFs based on the additive 
delay model \cite{lee2004technique} implemented in our in-house C-code. The additive delay 
model stipulates that the traveling times for the two signals on the two paths of the PUF 
to arrive at the arbiter are the summations of the gate delays incurred at all stages of the PUF. For the gate delays, we sampled from the Gaussian distribution with a mean of 300 and a deviation of 40 \cite{alkatheiri2017towards,aseeri2018XOR}. We created three 64-stage simulated FF~PUF instances, with different loop patterns for the three PUF instances. We generated 
five million CRPs for each simulated PUF instance.

With the CRPs ready, we carried out tests of the attack on a MacBook Pro with a 2.6 GHz 6-Core 
Intel Core i7 processor and a memory capacity of 16 GB. The neural network model for each 
simulated or silicon PUF is optimized for up to 300 epochs with an early stopping of 
5 patience. The experimental study used an 85-5-10 training-validation-testing split, 
where 85\% of CRP data were used for training, 5\% of data were used for validation, and 
10\% of data for testing the model.

When attacking each PUF instance, the number of CRPs 
used in the attack started small and gradually increasing until having reached a size 
(the size listed in the Column CRPs in \ref{tab6}), which resulted in successful attacks 
for all 30 PUF instances. The experimental results of attacks on 64-bit FF~PUFs without
defensive interface are listed in Table~\ref{tab6}. Results of attacks on interfaced 
FF~PUFs are listed in Table~\ref{tab7}, where the number of ghost bits ($m$) was chosen
to be the same as the number of feed-forward loops of the each FF PUF.

\def\arraystretch{1.1}
\begin{table}[]
\centering
\caption{Attacking 64-stage FF PUFs without Defensive Interfaces}
\label{tab6}
\begin{tabular}{|c|c|c|c|}
\hline
{\textbf{FF Loops}}&\textbf{CRPs}&\textbf{Avg. Accuracy}&\textbf{Training Time}\\ \hline
\textbf{4}         & 70K         & 98\%                 & 3 min                \\ \hline
\textbf{5}         & 180K        & 98\%                 & 7 min                \\ \hline
\textbf{6}         & 440K        & 97\%                 & 21 min               \\ \hline
\textbf{7}         & 520K        & 98\%                 & 31 min               \\ \hline
\textbf{8}         & 600K        & 98\%                 & 42 min               \\ \hline
\textbf{9}         & 770K        & 96\%                 & 63 min               \\ \hline
\textbf{10}        & 1.0M        & 93\%                 & 73 min               \\ \hline
\end{tabular}
\end{table}

\begin{table}[]
\centering
\caption{Attacking Defensively Interfaced 64-stage FF PUFs}
\label{tab7}
\begin{tabular}{|c|c|c|c|}
\hline
{\textbf{FF Loops}}&\textbf{CRPs}&\textbf{Avg. Accuracy}&\textbf{Training Time}\\ \hline
\textbf{4}         & 370K        & 94\%                 & 31 min            \\ \hline
\textbf{5}         & 4.5M        & Lower 60\%         & No convergence    \\ \hline
\textbf{6}         & 4.5M        & Upper 50\%         & No convergence    \\ \hline
\textbf{7}         & 4.5M        & Lower 60\%         & No convergence    \\ \hline
\textbf{8}         & 4.5M        & Lower 60\%         & No convergence    \\ \hline
\textbf{9}         & 4.5M        & Lower 60\%         & No convergence    \\ \hline
\textbf{10}        & 4.5M        & Mid ~ 50\%         & No convergence    \\ \hline
\end{tabular}
\vspace{1ex}
\end{table}



\section{Conclusion}

Many IoT devices are resource-constrained and demand security mechanisms implementable with 
low cost and operable with low energy. Leveraging integrated circuits' internal variability
as hardware fingerprints, physical unclonable functions (PUFs) have the potential as underlying 
primitives for implementing lightweight security protocols. The arbiter PUF and its variants
are highly lightweight in resource requirements, presenting themselves good candidates for resource-constrained IoT devices. But many of the PUFs have succumbed to machine learning
attacks. This paper introduces a defensive interface for the arbiter PUF and its variants. 
With the interface, experimental attack studies showed that the resilience of the tested PUFs
against machine learning attacks are substantially improved. The defensive interface incurs  
a low resource overhead, and hence can maintained in lightweightness of arbiter PUFs and
its variants. With the low resource-requirement and the capability of substantially improving
resistance to machine learning attack, the defensive interface is a promising candidate for
the protection of arbiter PUF variants.

\section{Acknowledgement}
The research was supported in part by the National Science
Foundation under Grant No. CNS-1526055.


\bibliographystyle{IEEEtran}
\bibliography{main.bib}

\end{document}